\newcommand{\version}{May 31, 2010}
\documentclass[11pt,a4paper,english]{article}

\usepackage[bindingoffset=0.3cm,textheight=22.5cm,hdivide={2.7cm,*,2.7cm}, vdivide={*,22cm,*}]{geometry}
\usepackage[dvips,bookmarksnumbered=true,breaklinks=true]{hyperref}

\usepackage{amsmath,amsfonts,amssymb,babel}
\usepackage{dsfont}
\let\mathbb=\mathds

\usepackage[numbers,square,comma,sort&compress]{natbib}

\newcommand{\startappendix}{\appendix}

\newcommand{\Appendix}[1]{
\vspace{3ex}
\pagebreak[2]
\refstepcounter{section}
\begin{flushleft}
{\Large\bf Appendix \thesection: #1}
\addcontentsline{toc}{section}{Appendix \thesection}
\end{flushleft}}

\newcounter{multieqs}

\newtheorem{theorem}{Theorem}
\newtheorem{lemma}[theorem]{Lemma}

\newcommand{\qed}{\nobreak \ifvmode \relax \else
      \ifdim\lastskip<1.5em \hskip-\lastskip
      \hskip1.5em plus0em minus0.5em \fi \nobreak
      \vrule height0.75em width0.5em depth0.25em\fi}

\newenvironment{proof}[1][Proof]{\begin{trivlist}
\item[\hskip \labelsep {\bfseries #1}]}{\qed\end{trivlist}}

\newcommand{\be}{\begin{equation}}
\newcommand{\ee}{\end{equation}}
\newcommand{\eq}[1]{(\ref{#1})}

\def\nn{\nonumber}
\def\bea{\begin{eqnarray}}
\def\eea{\end{eqnarray}}

\def\beqa{\begin{eqnarray}} 
\def\eeqa{\end{eqnarray}} 
\def\beq{\begin{equation}} 
\def\eeq{\end{equation}}

\def\Tr{{\rm Tr}}

\def\a{\alpha}          
\def\b{\beta}           
  \def\C{\Gamma}  
\def\d{\delta}    
\def\e{\epsilon}

\def\k{\kappa}
\def\l{\lambda} \def\L{\Lambda} 
\def\m{\mu}     \def\n{\nu}

\def\r{\rho}
\def\s{\sigma}  

\def\th{\theta}

\def\cJ{{\cal J}}  
\def\cM{{\cal M}} \def\cN{{\cal N}} \def\cO{{\cal O}}
\def\cP{{\cal P}}

\def\R{{\mathbb R}}
\def\C{{\mathbb C}}

\def\one{\mbox{1 \kern-.59em {\rm l}}}

\def\msu{\mathfrak{s}\mathfrak{u}}
\def\mmu{\mathfrak{u}}

\def\bit{\begin{itemize}}
\def\eit{\end{itemize}}

\def\({\left(}
\def\){\right)}

\def\d{\delta}

\def\pa{\partial} \def\del{\partial}

\def\uno{\mbox{1 \kern-.59em {\rm l}}}

\newcommand{\tr}{\mbox{tr}}

\def\Box{\square}

\def\bcomment#1{}

\newcommand{\uim}{UV/IR mixing}
\newcommand{\nc}{non-com\-mu\-ta\-tive}

\newcommand{\eqnref}[1]{Eqn.~(\ref{#1})}		
\newcommand{\inv}[1]{\frac{1}{#1}}				
\newcommand{\tinv}[1]{\tfrac{1}{#1}}
\newcommand{\lap}{\square_g}
\newcommand{\Lap}{\square_G}
\newcommand{\pb}[2]{\{#1,#2\}}						
\newcommand{\co}[2]{[#1,#2]}						
\newcommand{\aco}[2]{[#1,#2]_+}						
\newcommand{\intg}{\int\!d^4x\sqrt{g}\,}					

\renewcommand{\a}{\alpha}
\renewcommand{\b}{\beta}
\renewcommand{\d}{\delta}

\renewcommand{\th}{\theta}

\renewcommand{\l}{\lambda}

\renewcommand{\r}{\rho}
\newcommand{\vph}{\varphi}

\renewcommand{\L}{\Lambda}
\renewcommand{\Xi}{\Xi}

\title{\begin{flushright}
       \small{UWThPh-2010-3}
       \end{flushright}
\vspace{3em}
Curvature and Gravity Actions for Matrix Models}
\author{Daniel N. Blaschke\footnote{daniel.blaschke@univie.ac.at}~, Harold Steinacker\footnote{harold.steinacker@univie.ac.at}}

\date{\version}
\begin{document}
\maketitle

\begin{center}
\renewcommand{\thefootnote}{\fnsymbol{footnote}}
\textit{Faculty of Physics, University of Vienna\\
Boltzmanngasse 5, A-1090 Vienna (Austria)}
\vspace{0.5cm}
\end{center}%
\begin{abstract}

We show how gravitational actions, in particular the Einstein-Hilbert action, 
can be obtained from additional terms in Yang-Mills matrix models.
This is consistent with recent results on induced gravitational actions 
in these matrix models, realizing space-time as 4-dimensional
brane solutions. It opens up the possibility for a 
controlled non-perturbative
description of gravity through simple matrix models, with interesting
perspectives for the problem of vacuum energy. 
The relation with {\uim} and {\nc} gauge theory is discussed.

\end{abstract}

\newpage
\tableofcontents

\section{Background}\label{sec:background}
For a long time people have tried to combine the ideas of Quantum Mechanics and General Relativity 
in a consistent manner. In fact, such a combination strongly suggests a quantum structure of 
space-time itself near the Planck scale. While some aspects of such a quantum space-time can be seen 
in string theory or loop quantum gravity, a satisfactory understanding is still missing. 
A different approach to this problem has been discussed in recent years: Classical space-time 
is replaced by a quantized or {\nc} (NC) space, 
where the coordinate functions $x^\m$ are replaced by 
matrices resp. Hermitian operators $X^\m$ acting on a 
Hilbert space $\mathcal{H}$, which satisfy some non-trivial commutation relations
\begin{align}\label{eq:commutator}
\co{X^\m}{X^\n}&=i\th^{\m\n}\,.
\end{align}
While the simplest case of a Heisenberg algebra, i.e. with constant commutator $\th^{\m\n}$, 
leads to {\nc} field theories (cf.~\cite{Doplicher:1994tu,Douglas:2001ba,Szabo:2001kg,Rivasseau:2007a} 
for a review of the topic), a dynamical commutator seems essential in the context of gravity. 
At the semi-classical level, the commutation relations \eqref{eq:commutator} determine a 
Poisson structure $\th^{\m\n}$ on space-time, which is expected to be dynamical. 

The approach we would like to take is to consider matrix models of Yang-Mills type, 
which have already been shown to incorporate gravity, at least at the semi-classical 
level~\cite{Steinacker:2007dq,Grosse:2008xr,Klammer:2008df}. In this spirit, we follow 
Ref.~\cite{Steinacker:2008ri} by considering the matrix model action
\begin{align}\label{eq:basic-action-D-dim}
S_{YM}&=-\Tr\co{X^a}{X^b}\co{X^c}{X^d}\eta_{ac}\eta_{bd}\,,
\end{align}
where $\eta_{ac}$ is the (flat) metric of a $D$ dimensional embedding space 
(i.e. $a,b,c,d\in1,\ldots,D$). It can be purely Euclidean, or have one or more time-like directions. 
The ``covariant coordinates'' 
$X^a$ are Hermitian matrices, resp. operators acting on a separable Hilbert space $\mathcal{H}$. 
We denote the commutator of two coordinates as
\begin{align}
\co{X^a}{X^b}&=i \th^{ab}\,.
\end{align}
Furthermore, we consider for simplicity\footnote{However, this framework is not restricted to spaces with trivial topology, as can be seen e.g. by the example of a fuzzy sphere~\cite{Steinacker:2003sd}.} configurations where some of the $X^a$ are functions of
the remaining ones, because we are interested in $2n$ dimensional
{\nc} spaces $\mathcal{M}_\th^{2n}$.
We can then split the matrices resp. coordinates as
\begin{align}
X^a=\left(X^\m,\phi^i\right)\,,\qquad \m=1,\ldots,2n\,,\qquad i=1,\ldots,D-2n\,,
\end{align}
so that the $\phi^i(X) \sim \phi^i(x)$ define in the semi-classical limit
an embedding of a $2n$ dimensional submanifold
\be
\mathcal{M}^{2n}\hookrightarrow \R^D .
\ee 
Moreover, we can interpret
\begin{align}
\co{X^\m}{X^\n}&\sim i \th^{\m\n}(x)\,
\end{align}
in the semi-classical limit as a Poisson structure on $\mathcal{M}^{2n}$.
Thus we are considering quantized 
Poisson manifolds $(\cM^{2n},\theta^{\mu\nu})$, with quantized
embedding functions $X^a$.
Throughout this paper, $\sim$ denotes the 
semi-classical limit, where commutators are replaced by Poisson brackets.
We will assume that $\th^{\m\n}$ is non-degenerate, so that its 
inverse matrix $\th^{-1}_{\m\n}$ defines a symplectic form 
on $\mathcal{M}^{2n}$.
The sub-manifold $\cM^{2n}\subset\R^D$ 
is equipped with a non-trivial induced metric
\begin{align}
g_{\m\n}(x)=\pa_\m x^a \pa_\n x^b\eta_{ab}
=\eta_{\m\n}+\pa_\m \phi^i\pa_\n\phi^j\eta_{ij}\,,
\label{eq:def-induced-metric}
\end{align}
via pull-back of $\eta_{ab}$. 
Finally, we define the following quantities \cite{Steinacker:2008ri}:
\begin{align}\label{eq:notation}
G^{\m\n}&=e^{-\s}\th^{\m\r}\th^{\n\s}g_{\r\s}\,, 
      & \eta&=\inv{4}e^\s G^{\m\n}g_{\m\n}\,, \nonumber\\
\r&=\sqrt{\det{\th^{-1}_{\m\n}}}\,, 
       &  e^{-\s}&=\frac{\r}{\sqrt{\det{G_{\m\n}}}}\,.
\end{align}
The last relation gives a unique definition for $e^{-\sigma}$ provided $n>1$, which we assume.
Of particular interest is the special case where $2n=4$ and
\begin{align}
G^{\m\n}&=g^{\m\n} \qquad \rightarrow \qquad \eta=e^\s\,,
\end{align}
which corresponds to $\th^{\m\n}$ being self-dual with respect to the metric $g_{\m\n}$
(cf. \cite{Steinacker:2008ya}).

In order to understand the effective geometry of $\cM^{2n}$,
consider a test-particle on $\cM^{2n}$, 
modeled by a scalar field $\varphi$ for simplicity
(this could be e.g. an $\msu(k)$ component of $\phi^i$).
In order to preserve gauge invariance, 
the kinetic term must have the form
\begin{eqnarray}
S[\varphi] &\equiv& - \Tr [X^a,\varphi][X^b,\varphi] \eta_{ab} 
\sim \frac 1{(2\pi)^n}\, \int d^{2n} x\; \frac 1{|\theta^{\mu\nu}|^{1/2}}\,\,
e^a(\varphi)  e^b(\varphi) \eta_{ab} \nn\\
&=&   \frac 1{(2\pi)^n}\, \int d^{2n} x\; \frac 1{|\theta^{\mu\nu}|^{1/2}}\,
 \theta^{\mu\mu'}(x) \theta^{\nu\nu'}(x) g_{\mu\nu}
\partial_{\mu'}\varphi \partial_{\nu'}\varphi  \nn\\
 &=&\frac 1{(2\pi)^n}\, \int d^{2n} x\; 
|G_{\mu\nu}|^{1/2}\,G^{\mu\nu}(x)
 \partial_{\mu} \varphi \partial_{\nu} \varphi \,,
\label{covariant-action-scalar}
\end{eqnarray}
denoting the $D$ natural vector fields on $\cM^{2n}$ defined
by the matrix model as 
\begin{align}
e^a(f) := -i [X^a,f] \,\sim\, \theta^{\mu\nu}\partial_\mu x^a 
\partial_\nu f\, ,
\label{eq:def-vector-fields}
\end{align}
in the semi-classical limit where commutators are replaced by Poisson brackets. 
Therefore the kinetic term for $\varphi$ on  $\cM^{2n}$
is governed by the effective metric $G_{\mu\nu}(x)$, 
which depends on the Poisson tensor $\theta^{\mu\nu}(x)$ and the 
embedding metric $g_{\mu\nu}(x)$. In fact, 
the same metric also governs non-Abelian gauge fields 
and fermions  
in the matrix model (up to possible conformal factors), 
so that $G_{\mu\nu}$ {\em must} be interpreted as
gravitational metric. There is no need and no room 
for invoking any ``principles''. 
Since the embedding $\phi^i$ is dynamical, the model describes a dynamical
theory of gravity, realized on dynamically 
determined submanifolds of $\R^D$.

We furthermore note that 
\begin{align}
|G_{\mu\nu}(x)| = |g_{\mu\nu}(x)| , \qquad \mbox{2n=4}
\label{G-g-4D}
\end{align}
which means that in the 4-dimensional case, 
the Poisson tensor $\theta^{\mu\nu}$ does 
not enter the Riemannian volume at all. This 
turns out to stabilize flat space,
and is one of several reasons why 4 dimensions are special
in this framework.

\paragraph{Equations of motion}
The bare matrix model  \eqnref{eq:basic-action-D-dim} without matter
leads to the following e.o.m. for $X^c$:
\begin{align}\label{eq:basic-matrix-eom}
\co{X^a}{\co{X^b}{X^c}}\eta_{ab}&=0\,.
\end{align}
It was shown in Ref.~\cite{Steinacker:2008ri} that in the semi-classical limit, 
these equations can be brought into the covariant form using \eqnref{box-laplace-id}
\begin{subequations}
\begin{align}
\Lap \phi^i &= 0 ,   \label{eom-phi} \\
\Lap x^\mu &= 0 ,
\label{eom-X-harmonic-tree}
\end{align}
\end{subequations}
which imply
\begin{align}
\nabla^\m_G\left(e^\s\th^{-1}_{\m\n}\right)&=G_{\m\n}\th^{\m\r}e^{-\s}\pa_\r\left(e^\s\eta\right)\,.
\label{cons-theta}
\end{align}
Here $\nabla_G$ denotes the Levi-Civita connection with respect to the effective metric $G^{\mu\nu}$, 
and $\Lap$ denotes the corresponding Laplace-Beltrami operator. \eqnref{cons-theta} provides the 
relation between the non-commutativity $\theta^{\mu\nu}(x)$ and the metric $G^{\mu\nu}$.
Since it essentially has the form of covariant Maxwell equations coupled to an external current, 
it will have a unique solution for suitable ``boundary conditions''
\begin{align}
\theta_{\mu\nu}(x) \,\to
\, \bar \theta_{\mu\nu} = \text{const}
\qquad \mbox{for}\quad |x|\to\infty ,
\label{theta-asympt}
\end{align}
up to radiative contributions.

\paragraph{Ward identity}

For the matrix model \eqref{eq:basic-action-D-dim} one can derive the 
``energy-momentum tensor''
\begin{align}
T^{ab}&=\frac 12 \aco{\co{X^a}{X^c}}{\co{X^b}{X^{c'}}}\eta_{cc'}
-\inv{4}\eta^{ab}\co{X^c}{X^d}\co{X^{c'}}{X^{d'}}\eta_{cc'}\eta_{dd'}\,,
\label{eq:def-T}
\end{align}
whose conservation follows directly from the matrix equations of motion \eqref{eq:basic-matrix-eom} above:
\begin{align}\label{eq:basic-ward-id}
\co{X^a}{T^{a'b}}\eta_{aa'}=0\,.
\end{align}
In fact, this conservation constitutes a matrix Ward identity
(cf.~\cite{Steinacker:2008ri,Steinacker:2008ya}) corresponding to the 
infinitesimal transformation 
\begin{align}
\d X^a = \aco{X^b}{[X^a,\e_b]},
\label{dX}
\end{align}
for arbitrary matrices $\e_b = \e_b(X)$. \eqnref{dX} defines a 
measure-preserving infinitesimal transformation on the space of matrices which satisfies
\begin{align}
\d S = -8 \Tr \epsilon_b [X^a,T^{a'b}]\eta_{aa'}\,.
\end{align}
This implies \eq{eq:basic-ward-id} since the $\epsilon_b$ are arbitrary.

\section{Geometric considerations}\label{sec:geometry}

Since space-time is described in the matrix model as 4-dimensional
submanifold $\cM^4 \subset \R^D$, one should expect that both the
intrinsic as well as the extrinsic geometry of $\cM^4$ will play some role. 
In this section, we provide the necessary tools for an efficient
description of the geometry and the intrinsic (Riemannian) 
curvature of such branes; 
for a related discussion see e.g. \cite{Paston:2007qr}.
We restrict ourselves mostly to 4-dimensional configurations with
\be
G_{\mu\nu} = g_{\mu\nu} \,.
\ee
One can easily see\footnote{by going to local coordinates where $g_{\m\n}$ is diagonal at a point and $\th^{\m\n}$ has canonical form~\cite{Steinacker:2008ya}} 
that for 4-dimensional spaces $\cM^4$, this
is equivalent to the symplectic form
\be
\omega = \frac 12 \theta^{-1}_{\mu\nu} dx^\mu \wedge dx^\nu
\ee
being (anti-) self-dual\footnote{In the case of Minkowski signature, time-like
matrices $X^0$ should be anti-hermitian as explained in 
\cite{Steinacker:2008ya}. We then adopt the convention that 
$\varepsilon^{0123}$ is imaginary, so that $\star^2 = 1$.}, $\star \omega =\pm \omega$.
This imposes no significant restriction on the effective 
geometry $G_{\mu\nu} = g_{\mu\nu}$, since
such (anti)self-dual $\omega$ can essentially always be found
for a given metric (assuming e.g. that $\cM^4$ is globally hyperbolic).
In that case $\eta = e^\sigma$ (cf. \eqref{eq:notation}), and \eqnref{cons-theta} reduces to 
\be
\nabla^\mu \theta^{-1}_{\mu\nu} = 0 ,
\ee
which is satisfied identically for self-dual $\omega$.
We can drop the subscripts $g$ or $G$ to the covariant derivatives
from now on.

Under this assumption,
the conserved ``tensor'' $T^{ab}$ acquires a simple geometrical meaning
in the semi-classical limit: it essentially becomes 
the projector on the normal space 
$N_p\cM^4 = (T_p\cM^4)^\perp$. More precisely,
\begin{align}
T^{ab}&\sim e^\s\,  \cP_N^{ab},\nonumber\\
\cP_T^{ab} &=g^{\m\n}\pa_\m x^a\pa_\n x^b\,, \qquad
\cP_N^{ab} = \eta^{ab} - \cP_T^{ab}
\,,
\label{eq:T-prop}
\end{align}
where $\cP_{N,T}$ are the projectors on the normal resp. tangential space
at $p\in \cM^4$. This means that
\begin{align}
\cP_T^{ab}\eta_{bc}\pa_\m x^c&=\pa_\m x^a\,,\qquad
\cP_N^{ab}\eta_{bc}\pa_\m x^c =0 
\,, \nn \\
\cP_T^2  &= \cP_T, \qquad\qquad\qquad \cP_N^2  = \cP_N
\,,
\label{projector-prop}
\end{align}
which is easy to verify.
Here
\begin{align}
\del_\mu \equiv (\del_\mu x^a)_{a=1,2,...,D} \quad\in T_p\cM^4\,\,\subset \R^D
\end{align}
is interpreted as tangent vector field defined by some coordinate system,
represented in $D$-component notation.
This allows to write down covariant derivatives
 $\nabla \equiv\nabla_g$
with respect to the embedding metric. For example, the covariant derivative of
the vector field $\del_\mu$ defined by some coordinate system 
(i.e. $V = V^\nu_{(\mu)}\del_\nu$ with $V^\nu_{(\mu)} = \delta^{\nu}_\mu$) is 
\begin{eqnarray}
(\nabla_\mu \del_\nu)^a &=& \cP_T^{ab}\eta_{bc} \del_\mu\del_\nu x^c
= \left(g^{\l\sigma} \del_\sigma x^b
\eta_{bc}\del_\mu\del_\nu x^c \right) \del_\l x^{a}
= \Gamma^\l_{\mu\nu} (\del_\l)^a\,,
\label{cov-der-VF}
\end{eqnarray}
where 
\begin{align}
\Gamma^\l_{\mu\nu} = g^{\l\sigma}\del_\sigma x^b
\eta_{bc}\del_\mu\del_\nu x^c\,
\label{christoffel-explicit}
\end{align}
is the Christoffel symbol w.r.t. $g_{\mu\nu}$. 
From this it is easy to recover the
standard formula in terms of $g_{\mu\nu}$.
Notice that using the $D$-dimensional Poincare symmetry, we can choose
for any given point $p\in \cM^4$ the matrix
coordinates $x^a = (x^\mu,\phi^i)$ such that $\del_\mu \phi^i|_p =
0$. These are called matrix normal coordinates, which satisfy
$\left.\Gamma^\l_{\mu\nu}\right|_p =0$.
Similarly, the second covariant derivative of the scalar fields
defined by the matrices  $x^a: \cM^4 \to \R^D$
(i.e. the second fundamental form) is given by\footnote{Notice the difference 
to \eq{cov-der-VF}, where $\del_\mu$ is
interpreted as a vector field.}
\be
\nabla_\mu\nabla_\nu x^a = \cP_N^{ab}\eta_{bc} \del_\mu\del_\nu x^c
= (\del_\mu\del_\nu  - \Gamma^\rho_{\mu\nu}\del_\rho) x^a .
\label{eq:projector-properties2}
\ee
It immediately follows that
\be
\nabla_\mu x^a \nabla_\nu\nabla_\rho x_a = 0 ,
\label{3-nabla-vanish}
\ee
which can also be seen from $\nabla g = 0$ or by going to normal 
embedding coordinates. Here and in the following we adopt the convention that 
Latin indices of $X^a \sim x^a$ are raised or lowered with 
the constant $D$ - dimensional background metric $\eta_{ab}$.
It follows that
\be
\cP^{ab}_N\nabla_\mu\nabla_\nu x_b = \nabla_\mu\nabla_\nu x^a
\label{nablanabla-perp}
\ee
which will be used below.

We can now write down the Riemann curvature tensor in a 
useful form. Consider
\bea
(-{R_{\nu\mu\l}}^\kappa \del_\kappa)^a &=&
([\nabla_\nu,\nabla_\mu]\del_\l)^a \nn\\
&=&  (\cP_T^{ab}\del_\nu \cP_T^{b'c} \del_\mu - \cP_T^{ab}\del_\mu \cP_T^{b'c} \del_\nu)
\del_\l x_c \eta_{bb'}  \nn\\
&=& \cP_T^{ab}(\del_\nu \cP_T^{b'c} \del_\mu - \del_\mu \cP_T^{b'c} \del_\nu)
\del_\l x_c \eta_{bb'}  \nn\\
&=& \del_\kappa x^a \Big(\del^\kappa x^b
 \del_\nu \cP_T^{b'c} \del_\mu \del_\l x_c
- \del^\kappa x^b\del_\mu \cP_T^{b'c} \del_\nu \del_\l x_c\Big) \eta_{bb'}
\,,
\eea
hence
\bea
{R_{\nu\mu\l\kappa}}
&=& -\del_\nu \cP_T^{bc}\del_\kappa x_b  \del_\mu \del_\l x_c
+ \del_\mu\cP_T^{bc}\del_\kappa x_b  \del_\nu \del_\l x_c 
\,, \nn\\
R_{\nu\l} &=& {R_{\nu\mu\l}}^{\mu} 
= -\del_\nu \cP_T^{bc}\del^\mu x_b  \del_\mu \del_\l x_c
+ \del_\mu\cP_T^{bc}\del^\mu x_b  \del_\nu \del_\l x_c 
\,, \nn\\
R &=& R_{\nu\l} g^{\nu\l}
= \del_\nu \cP_T^{bc} \Big(-\del^\mu x_b  \del_\mu \del_\l x_c g^{\l\nu}
+ \del^\nu x_b \del_\mu \del_\l x_c g^{\l\mu} \Big).
\eea
Using \eq{projector-prop} 
and noting
$\del_\nu \cP_T^{bc}\del_\kappa x_b 
 = \del_\nu (\cP_T^{bc}\del_\kappa x_b) - \cP_T^{bc}\del_\nu\del_\kappa x_b$ 
this can be written as 
\bea
R_{\nu\mu\l\kappa} 
 &=& -\cP_N^{ab}\, (\del_{\k}\del_{\nu}x_a \del_{\l}\del_{\mu}x_b
- \del_{\k}\del_\mu x_a\del_{\nu}\del_{\l}x_b )\nn\\
&=& -\nabla_{\k}\nabla_{\nu}x^a \nabla_{\l}\nabla_{\mu}x_a
+ \nabla_{\k}\nabla_\mu x^a\nabla_{\nu}\nabla_{\l}x_a 
\label{R-proj-full}
\eea
which is nothing else but the Gauss-Codazzi theorem. In particular,
the Ricci scalar is given by
\bea
R &=& -\cP_N^{ab}\, (\del_{\k}\del_{\nu}x_a \del_{\l}\del_{\mu}x_b
- \del_{\k}\del_\mu x_a\del_{\nu}\del_{\l}x_b ) g^{\k\mu}g^{\l\nu}\nn\\
 &=& -\nabla_\mu\nabla_{\nu}x^a \nabla^\mu\nabla^{\nu} x_a
 + \lap x^a \lap x_a \,.
\label{R-nabla}
\eea
Finally, we note that the conservation law \eq{eq:basic-ward-id}
in the semi-classical limit reduces to 
\bea
0 &=& \theta^{\mu\nu}\del_\mu x_b \del_\nu T^{bc} 
= \theta^{\mu\nu}\del_\nu \(\del_\mu x_b\, T^{bc} \)
\,.
\label{eq:ward-id_semiclass}
\eea
This holds as long as $T^{bc}$
is a projector on the normal bundle (see Eqns.~\eqref{eq:T-prop} and \eqref{projector-prop}), which follows from $g_{\mu\nu} = G_{\mu\nu}$.

\section{Extensions to the matrix model action}\label{sec:extensions}

In this section we would like to discuss some possible extensions to the matrix model action \eqref{eq:basic-action-D-dim}. In particular, we will find terms which depend only on the 
intrinsic geometry of $\cM^4 \subset \R^D$, including essentially the 
Einstein-Hilbert action as well as a term coupling the 
Riemann tensor to the Poisson tensor. This allows to realize 
Einstein gravity (in a slightly modified form) 
and its quantization through matrix models.
The terms obtained are in agreement with the 
one-loop effective action for the gravitational sector of the Yang-Mills
matrix model \cite{Steinacker:2008ri,Klammer:2009dj}.
This shows that the quantization of the model can be addressed  
both from the geometrical point of view as well as from the 
matrix model point of view. 
It opens up the possibility for a controlled
non-perturbative quantization of the matrix model
by adding suitable counter terms. 

Throughout this section we consider 4-dimensional $\cM^4 \subset \R^D$
with self-dual $\theta^{-1}_{\mu\nu}$, i.e. $G_{\m\n}=g_{\m\n}$.

\subsection{Order 6 terms}\label{sec:order6}

We first note the following identities:
\begin{align}
& \Tr\, \Box X^a \Box X_{a} =  \Tr \Big( \tinv{2} [X^c,[X^a,X^b]] [X_c,[X_a,X_b]]
  - 2\co{X^a}{X^c}\co{X_c}{X^b}\co{X_a}{X_b}\Big) \,,\nn\\
& \Tr[X^a,[X^b,X^c]] [X_c,[X_a,X_b]] =
-\tinv{2} \Tr[X^c,[X^a,X^b]] [X_c,[X_a,X_b]]\,,
\end{align}
using the abbreviation
\begin{align}\label{eq:def-Box}
\Box X^a\equiv \co{X^b}{\co{X_b}{X^a}}\,.
\end{align}
This leaves the following independent terms:
\begin{align}
S_6&= \Tr \left(\a \Box X^a \Box X_a + \frac{\b}{2} [X^c,[X^a,X^b]] [X_c,[X_a,X_b]]\right)
\,.
\end{align}
It turns out that these terms have a nice geometrical meaning in the 
semi-classical limit. As shown in Appendix \ref{sec:appendix}, one finds
\begin{align}
S_6&\sim \frac{\a+\b}{(2\pi)^2} \intg  e^\s \lap x^a \lap x_a
+  \frac{\b}{(2\pi)^2}\intg  \bigg(\frac 12\theta^{\mu\rho} \theta^{\eta\a} R_{\mu \rho\eta\a} 
- 2 e^\s R 
+ 2 e^\s \del^{\mu}\s \del_\mu\s \!\bigg) 
\label{S6-geom}
\end{align}
using \eq{R-nabla}. The first term depends on the ``extrinsic''
geometry, i.e. the embedding $\cM^4 \subset \R^D$, and vanishes for
harmonic embeddings where  $\lap x^a =0$. 
However, for $\a+\b=0$ the result is purely tensorial and
intrinsic, 
independent of the particular embedding of $\cM^4 \subset \R^D$. This is
an essential feature of General Relativity. We will see that 
such terms are also induced at one loop when coupling fermions 
to the matrix model~\cite{Klammer:2009dj}, hence the above terms
can be used to cancel unwanted terms in the quantum effective action.
The Einstein-Hilbert action is obtained from similar
higher-order terms, as we show next.

\subsection{Higher order terms}\label{sec:higher-order}

There are other terms in the matrix model which involve up to 
4 derivatives in the semi-classical limit. Rather than giving an 
exhaustive list we only discuss some terms of particular interest here.

\paragraph{Order $X^{10}$ terms.}

Consider the following order 10 terms in the 
semi-classical limit:
\begin{subequations}
\begin{align}
S_{10}&=(2\pi)^2\Tr\left(\co{X^a}{T^{bc}}\co{X_a}{T_{bc}}+2T^{ab}\Box X_b \Box X_c\right)\nn\\
&\sim \intg \left[(D-2n)e^{\s}\lap e^\s+2e^{2\s}R\right]\, \label{eq:S10} \\
\tilde S_{10}&=(2\pi)^2\Tr \co{X^a}{X^b}\co{X_a}{X_b}\Box\left(\co{X^c}{X^d}\co{X_c}{X_d}\right)\nn\\
&\sim -16\intg e^\s\lap e^\s\,,
\label{eq:S10-tilde}
\end{align}
\end{subequations}
where once more $D$ is the dimension of the embedding space, 
and $2n=4$ denotes the dimension of the {\nc} subspace.

\begin{proof}
The semi-classical limit of the first term of \eqref{eq:S10} is given by
\begin{align}
(2\pi)^2\Tr\co{X^a}{T^{bc}}\co{X_a}{T_{bc}}&\sim 
-\intg 
\del_\mu (e^\sigma P_N^{bc}) 
\del^\mu(e^\sigma(\eta_{bc} - \del^\nu x_b \del_\nu  x_c)) \nn\\
&=  -\intg 
\nabla_\mu (e^\sigma P_N^{bc}) 
\nabla^\mu(e^\sigma\eta_{bc} - e^\sigma\del^\nu x_b \del_\nu  x_c))
\nn\\
&= \intg e^\sigma P_N^{bc} \Big(
\eta_{bc} \nabla_\mu\nabla^\mu e^\sigma
- \nabla_\mu\nabla^\mu(e^\sigma\del^\nu x_b \del_\nu x_c) \Big)\nn\\
&= \intg e^\sigma \Big(
(D-2n) \lap e^\sigma
-2 P_N^{bc}(e^\sigma\nabla^\mu\del^\nu x_b \nabla_\mu\del_\nu x_c) \Big)\nn\\
&= \intg e^\sigma \Big(
 (D-4) \lap e^\sigma
- 2 e^\sigma\nabla^\mu\del^\nu x^a \nabla_\mu\del_\nu x_a \Big)
\end{align}
using the properties \eqref{projector-prop} and \eqref{eq:projector-properties2} of the projector $\cP_N^{bc}$.\\
The second term of \eqnref{eq:S10} semi-classically is
\begin{align}
(2\pi)^2\Tr\, T^{ab}\Box X_b \Box X_c&\sim\intg e^{2\s}\cP_N^{bc}\lap x_b \lap x_c\nn\\
&=\intg e^{2\s}\lap x^c \lap x_c
\end{align}
and using \eqref{R-nabla} one finally arrives at \eqnref{eq:S10}.
The second identity \eq{eq:S10-tilde} simply follows from
\begin{align}
e^\s\Big|_{G=g}=\eta=\inv{4}\th^{\m\n}\th^{\r\s}g_{\m\r}g_{\n\s}=\inv{4}\pb{x^a}{x^b}\pb{x_a}{x_b}\,,
\end{align}
where $\pb{x^a}{x^b}$ denotes the Poisson bracket.
\end{proof}
This means that the matrix model action\footnote{In fact, it is easy to see that the equality of the second and third line of \eqref{E-H-action} holds even without the trace, resp. the integral.} 
\begin{align}
S_{\text{E-H}}&=\Tr\left(\co{X^a}{T^{bc}}\co{X_a}{T_{bc}}+2T^{ab}\Box X_b \Box X_c
 + \tfrac{D-4}{16} \co{X^a}{X^b}\co{X_a}{X_b}
\Box(\co{X^c}{X^d}\co{X_c}{X_d})\right)\nn\\
&= \Tr\left(2T^{ab}\Box X_a \Box X_b - T^{ab}\Box H_{ab}\right)\nn\\
&\sim \frac{2}{(2\pi)^2} \intg e^{2\s}R 
\,,\label{E-H-action}
\end{align}
where (cp. \eqref{eq:def-T})
\begin{align}
H^{ab}&=\inv{2}\aco{\co{X^a}{X^c}}{\co{X^b}{X_c}}
\,,
\end{align}
reduces in the
semi-classical limit to the Einstein-Hilbert action, 
with an additional factor $e^{2\sigma}$ which introduces the required scale.
After introducing an explicit dimensionful parameter $\L_0$ of 
dimension length$^{-1}$ (so that $X^a\sim x^a$ acquires the 
appropriate dimension of length) 
to the matrix model, we can identify
the gravitational constant arising from this term as
\be
\L_{\rm planck}^2 = G = \frac{\L_0^{10}}{\L_{NC}^8}
\label{Grav-const}
\ee
recalling that $e^{\sigma} = \L_{NC}^{-4}$ sets the non-commutativity 
scale. It is thus reasonable to set $\L_{NC} \sim \L_0 \sim \L_{\rm  planck}$.
However, this should be taken with some caution
since quantum effects will play an important role,
as indicated below.

It is remarkable that one obtains in this way an action which 
depends only on the intrinsic geometry of $\cM^4 \subset \R^D$.
In particular every Ricci-flat manifold can be obtained as a solution 
of the matrix model with the term \eq{E-H-action}, for
selfdual $\omega$. Notice that self-dual $\omega$ is then indeed a solution,
since $\intg \d e^{2\sigma} R = 0$.
As an example, we will present an explicit realization of the Schwarzschild-solution
in \cite{workinprogress}.

\paragraph{Order  $X^{8}$ terms}
The simplest term of order $X^8$ is given by
\be
S_8 = \frac{(2\pi)^2}4\Tr([X^a,X^b][X_b,X_c][X^c,X^d][X_d,X_a])
\,.
\ee
The semi-classical limit of this term is obtained easily 
using \eqref{eq:def-vector-fields}, \eqref{eq:def-induced-metric} and \eqref{eq:notation} for the self-dual case, i.e. $G=g$ and hence $\eta=e^\s$:
\begin{align}
S_8 &\sim \intg\, e^{\sigma} \,.
\end{align}
A preliminary analysis suggests that the only other non-vanishing 
terms of order $X^8$ lead to higher-order derivative terms, such as 
$\Tr [X,[X,\theta]][X,[X,\theta]]$ where $\theta$ stands for $[X,X]$.
Such higher-derivative terms are typically suppressed 
at low energies and should be studied elsewhere.

\subsection{Potentials}\label{sec:potentials}

We now consider the possibility to add explicit "potential" terms 
to the matrix model which break the 
translational invariance
$X^a \to X^a + c^a \one$, but preserve the $D$-dimensional rotational invariance
of the matrix model. The 
simplest such extension to the matrix model is a ``mass'' like term of the form
\begin{align}\label{eq:mass-term}
S_m &= (2\pi)^2 m^2 \Tr X^a X_a
\, \sim m^2\intg e^{-\s} x^a x_{a} .
\end{align}
Similarly, one could also add higher powers of $X^a X_a \sim x^2$ to the action, for example
\begin{align}
S_{V4}&=(2\pi)^2\m_4\Tr\left(X^aX^bX^cX^d\left(\eta_{ab}\eta_{cd}+\eta_{ac}\eta_{bd}\right)\right)
\, \sim 2\m_4\intg e^{-\s} (x^2)^2 .
\label{eq:S-V4}
\end{align}
Now consider the equation of motion in
the presence of such potential terms. 
For example, upon adding a ``mass'' term $\Tr V(X)$ with $V(X) = m^2 X^a X_a$,
the e.o.m. of the matrix model becomes
\be
\Box X^a = \frac 12 m^2 X^a .
\ee
Using \eq{3-nabla-vanish}, this implies  semi-classically that 
\be
0 = \del_\mu x^a \Box_G x^a = \frac 12 m^2 e^{-\sigma} \del_\mu x^a x^a 
= \frac 14 e^{-\sigma} \del_\mu V(x)
\label{V-const}
\ee
which holds also for more general potentials  $V(X) \sim V(x)$.
This means that $V(x) = \text{const.}$, so that $\cM^4 \subset \R^D$
must be a sub-manifold of the equi-potential hypersurface.
This is well-known in the examples of fuzzy spaces such as
$S^2_N$ or $\C P^2$ \cite{CarowWatamura:1998jn,Steinacker:2003sd,Grosse:2004wm} where 
$X^a X_a = \text{const.} \one$.
As a consequence, the tangential
conservation law \eqref{eq:basic-ward-id}, which a priori is modified and reads
\bea
\co{X_a}{T^{ab}+\tfrac{m^2}{4}\aco{X^a}{X^b}} &=&0\,
\label{eq:ward-id_mass}  \\
\co{X_a}{T^{ab}} &=& -\frac{m^2}{4}[X_a X^a,X^b] \nn\\
&\sim & \frac{m^2}4 i\th^{\m\n}\pa_\m x^2\pa_\n x^b = 0 , \nn
\eea
is in fact unchanged and holds also in the presence of a potential,
since $\del_\mu V(x) = 0$ on $\cM^4$.
Therefore self-dual $\omega$ with $G_{\m\n}=g_{\m\n}$ supplemented with the additional 
condition $x^2=C=\text{const.}$ fulfill the modified (semi-classical) Ward identity 
\eqref{eq:ward-id_mass} --- 
and also the tangential conservation law \eqnref{eq:ward-id_semiclass}. 

We can easily extend the analysis of 
the potential terms to include the next-to-leading order (n.l.o.) 
corrections in $\theta^{\mu\nu}$.
To do this we replace the matrix product with the Groenewold-Moyal star product
(cf. \cite{Douglas:2001ba,Szabo:2001kg}) in Darboux coordinates, 
where $\th^{\m\n}$ is constant. Then
the semi-classical limit including n.l.o. corrections  
of the mass term  reads
\begin{align}
S_m &\sim m^2\intg e^{-\s}\Big(x^a x_{a} - \inv{8}\th^{\m\n}\th^{\m'\n'}\pa_\m\pa_{\m'} x^a
\pa_\n\pa_{\n'} x_{a}\Big)
\nonumber\\
&\sim m^2\intg e^{-\s}x^a x_{a}
\,,
\end{align}
where the last line follows from partial integration, 
noting that $\r=\sqrt{g}e^{-\s}$ is constant in Darboux coordinates. 
Similarly, the quartic term \eq{eq:S-V4} 
in the semi-classical limit including n.l.o. corrections reads
\begin{align}
S_{V4}&\sim \m_4\intg e^{-\s}\Big(2(x^2)^2-\inv{4}\th^{\m\n}\th^{\m'\n'}\left(x^2\pa_\m\pa_{\m'} x^a
\pa_\n\pa_{\n'} x^b\eta_{ab}+\inv{2}\pa_\m\pa_{\m'} x^2
\pa_\n\pa_{\n'} x^2\right)\nn\\
&\qquad\qquad\qquad -\frac{3}{8}\th^{\m\n}\th^{\m'\n'}\left(\tinv{2}\pa_\m\pa_\r x^2-g_{\m\r}\right)\left(\tinv{2}\pa_\n\pa_\s x^2-g_{\n\s}\right)\Big)\nn\\
&\sim \m_4\intg \bigg[2e^{-\s}(x^2)^2-\inv{4}e^{-\s}\th^{\m\n}\th^{\m'\n'}\left(x^2\pa_\m\pa_{\m'} x^a
\pa_\n\pa_{\n'} x^b\eta_{ab}+\frac{7}{8}\pa_\m\pa_{\m'} x^2
\pa_\n\pa_{\n'} x^2\right)\nn\\
&\qquad\qquad\qquad +\frac{3}{8}g^{\m\m'}\pa_\m\pa_{\m'} x^2-\frac{3}{8}\bigg]\,,\\
& \sim 2\m_4\intg e^{-\s}(x^2)^2
\end{align}
where $x^2\equiv x^ax^b\eta_{ab}$, using again $x^2 = \text{const.}$ on
solutions of the e.o.m.

However, some of the $O(X^6)$ terms considered above might 
modify the equation which determines $\theta^{\mu\nu}$, which should
be studied elsewhere.

\section{Non-Abelian sector}
\label{sec:nonabelian}

In this section we briefly discuss the relevance of the additional terms under
consideration of the non-Abelian sector of the model,
which arises on backgrounds corresponding to $n$ coinciding branes.
In order to avoid notational conflicts, 
we denote the basic matrices 
with $Y^a$ in this section, governed by 
the same matrix model as above
\be
S_{YM} = - \frac{\L_0^4}{4 g^2}\, \Tr [Y^a,Y^b] [Y^{a'},Y^{b'}] 
\eta_{aa'}\eta_{bb'} \,,
\label{YM-MM-nonabel}
\ee
but for a matrix background of the form
\be
Y^a = \left(\begin{array}{l}Y^\mu \\ Y^i \end{array}\right) = 
\left\{\begin{array}{ll}  X^\mu\otimes \one_n,  & \quad a=\mu = 1,2,..., 2n,   \\ 
\phi^i \otimes \one_n, & 
\quad a = 2n+i, \,\, i = 1, ..., D-2n . \end{array} \right.
\ee
Here we also introduce a dimensionful scale 
parameter $\L_0$ of dimension $length^{-1}$, so that
$X^a \sim x^a$ acquire the appropriate dimension of length.
As shown in \cite{Steinacker:2008ya}, the fluctuations of the 
tangential resp. transversal $\msu(n)$-components
of $Y^a$ lead to $\msu(n)$-valued gauge fields resp. scalar fields
coupled to $G_{\mu\nu}$. 
The Yang-Mills matrix model \eq{YM-MM-nonabel} then describes non-Abelian gauge theory
coupled to gravity, 
with effective gauge coupling ``constant'' 
\be
\frac 1{g_{YM}^2} = \frac{\L_0^4 e^\sigma}{g^2}
\,,
\ee
which reduces to $g_{YM}^2 \sim g^2$ 
assuming $\L_0 \sim \L_{NC}$ as discussed above.
Therefore any of the additional terms in the 
matrix model action discussed above also lead to additional
terms for the non-Abelian gauge fields. The form of
these terms is strongly restricted by gauge invariance.
As usual, any higher-order terms in the field strength must be 
suppressed on dimensional grounds 
by the non-commutativity scale $\frac 1{\Lambda_{NC}}$
(which arises from $\theta^{\mu\nu}$ or through the scalar field 
$e^\sigma$),
and are therefore irrelevant at low energies.
However, they may in general also contain explicit curvature terms.
For example, the order 6 terms are expected to contain terms of type
\begin{align}
S_6(F)\sim \frac{\L^6_0}{\L_{NC}^8}\intg \tr\Big(c_1 F^{\m\n}F^{\r\s}R_{\m\n\r\s}
+c_2 DFDF+c_3 F\co{F}{F}\Big)\,.
\label{S6-gauge}
\end{align}
In particular, the first of these possible terms explicitly depends on
the Riemann curvature and has a structure similar to the non-standard
term $\th\th R$ which already appeared in \eqnref{S6-geom}, 
where $F\leftrightarrow\th$. Such terms should be expected anyway
in the quantum effective action. Similarly,
the $O(X^{10})$ terms \eq{eq:S10} leading to the Einstein-Hilbert action 
are expected to contain the gauge structure
\be
S_{10}(F) \sim \frac{\L_0^{10}}{\L_{NC}^{16}}  \intg DF F DF F 
\label{S-10-scale}
\ee
as well as terms which are lower-order in $F$ such as \eq{S6-gauge}.

We close this section with a comment on the mass term \eq{eq:mass-term}.
At first sight, it may appear that it would
lead to a mass term for the non-Abelian gauge fields,
which would be in conflict with gauge invariance. However, this is
of course not the case, as is well-known in the examples of 
fuzzy spaces \cite{CarowWatamura:1998jn,Steinacker:2003sd,Grosse:2004wm}. 
A careful derivation of its effect on the $\msu(n)$ components
would require to use the 2nd order Seiberg-Witten 
map, which we will not carry through here.

\section{Quantization and one-loop effective action}

In this final section we briefly discuss the quantization of the
matrix model 
\begin{align} 
S_{\Psi}&=-\Tr\left(\inv{4}\co{X^a}{X^b}\co{X_a}{X_b}+\inv{2}\bar\Psi\gamma_a\co{X^a}{\Psi}\right) 
\label{action-fermions}
\end{align} 
and the significance of the additional terms which we have introduced above, 
restricting ourselves essentially to one loop. 

Several different points of view can be taken.
First, one can use the geometrical interpretation of the above model 
as an action for matter and fields on 
branes with non-trivial geometry $g_{\mu\nu}$, following 
\cite{Steinacker:2007dq,Grosse:2008xr}.
The (one-loop) induced action due to 
integrating out matter and fields can then be obtained from the 
standard heat kernel expansion on such a background, using
well-known Seeley-de Wit coefficients.
This leads to an induced Einstein-Hilbert
action, as well as additional ``non-standard'' terms. 
Assuming an effective UV-cutoff\footnote{This could arise 
either by adding an explicit UV-cutoff term such as $\Tr\, \Box X^a\Box X_a$,
or in the IKKT model \cite{Ishibashi:1996xs} upon adding soft SUSY breaking terms 
which may lead e.g. to compactification on 
spontaneously generated $S^2_N$ \cite{Aschieri:2006uw}.} 
$\Lambda$,  the contribution
due to scalar fields $\vph$ in the matrix model
(which arise as part of the $\msu(n)$ sector in the matrix model
in backgrounds of the form $X^a\otimes \one_n$) has the expected form
\cite{Steinacker:2008ri}
\begin{align}
\Gamma_\vph=\frac{1}{16\pi^2}\intg &\left(-2\L^4-\frac 16\L^2 R[g]\right)
\,.
\label{induced-grav-scalars}
\end{align}
The induced action due to fermions is more complicated, because the 
matrix Dirac operator in \eq{action-fermions} leads to a non-standard 
spin connection on general $\cM^4\subset \R^D$. 
For geometries with $g_{\mu\nu} = G_{\mu\nu}$,
the result can be written as~\cite{Klammer:2009dj}
\begin{align}
\Gamma_\Psi=\frac{k}{16\pi^2}\intg &\bigg[4\L^4
+\L^2\left(-\inv{3}R
+\inv{4} \del^\mu\s\del_\mu\s
+\inv{8}e^{-\s}\th^{\m\n}\th^{\r\s}R_{\m\n\r\s}
+\inv{4}\lap x^a\lap x_a\right)\nn\\
&\; +\cO(\log\L)\bigg]\,,
\end{align}
where $k$ is the number of components of $D$-dimensional 
Dirac fermions.
Remarkably, the terms of order $\L^2$ 
essentially coincide with the semi-classical limit of the
additional matrix model terms considered in 
Sections~\ref{sec:order6}--\ref{sec:higher-order} (up to
powers of $e^\s$ which provide the required scale
as in \eq{S-10-scale}).
Indeed, it should be possible to perform the quantization directly 
within the matrix model, leading to quantum corrections to the 
effective action within the framework of matrix models.
This should lead to a quantum effective action given by 
additional terms in the matrix model, such as
the terms studied above. We have therefore identified 
the corresponding terms in such an effective matrix model,
consistent with the semi-classical computation.
This also provides an indirect check for the results in 
\cite{Klammer:2009dj}.

Furthermore, having the above matrix model terms at our disposal,
we can use them as counter terms in order to cancel 
unwanted terms in the effective action 
such as $R\theta\theta$  or the 
``extrinsic'' term $\lap x^a \lap x^a$. One can then indeed
adjust the model such that the gravitational action 
reduces essentially to the 
Einstein-Hilbert action, plus higher-order 
terms which are suppressed by $\frac 1{\L}$.
Therefore the matrix model can be used to realize and to 
quantize general relativity, or some very closely related 
gravity theory.  

On the other hand, introducing such counter terms by hand will in
general spoil the good renormalization properties of the 
``pure'' Yang-Mills matrix model, which equivalently can be 
viewed as {\nc} gauge theory on $\R^4_\theta$.
In particular, the IKKT model \cite{Ishibashi:1996xs} (possibly 
with soft SUSY breaking terms such as a mass term) 
corresponds to $\cN=4$ NC SYM on $\R^4_\theta$, and is expected 
to be finite \cite{Jack:2001cr,Matusis:2000jf}.
Here the above terms should preferably be used only 
for intermediate steps, e.g. to introduce a 
controlled UV cutoff which should be removed in the end.
It is  quite conceivable that a realistic gravity action 
arises purely from the finite, induced gravitational terms 
below e.g. the $\cN=4$ breaking
scale. At this point, we should
briefly discuss the important aspect of 
UV/IR mixing NC gauge theories, and its relevance in the present context.

\paragraph{$\mmu(1)$ sector and UV/IR mixing}

It is well known in NC field theory that 
the renormalization of the $\mmu(1)$ and the $\msu(n)$ sectors
differ drastically at low energies\footnote{Even though this
  appears to break the full gauge invariance of the matrix model,
  this is not the case: the $U(1)$ invariance is simply transmuted
  into an invariance under symplectomorphisms.} due to UV/IR mixing 
\cite{Minwalla:1999px,Armoni:2000xr}: 
the $\mmu(1)$ sector diverges 
in the IR if the UV cutoff is removed. 
This is usually perceived as a problem but 
is in fact very welcome here and
consistent with the identification of the $\mmu(1)$
sector in terms of gravity. 
A careful analysis from the point of view of emergent gravity 
\cite{Grosse:2008xr} 
shows that the IR divergence is precisely due to the induced 
gravitational action in the $\mmu(1)$ sector. 
For example, the $\L^4$ divergence of the cosmological constant 
in \eq{induced-grav-scalars}
essentially arises from the IR limit of
the {\em effective cutoff} 
\bea
\L_{eff}^4(p) &=&\frac 1{(\frac 1{\Lambda^2} + \frac 14 \frac{p^2}{\L_{NC}^4})^2}
\sim \L^4 + ...
\label{lambda-eff}
\eea
in the quantum effective action. 
$\L_{eff}^4(p)$ is in fact of order $O(1)$ at $p\approx\L_{NC}$,
where $p$ is the momentum scale of the gravitational action.
Hence the gravitational $\mmu(1)$ sector
scales differently under renormalization than the non-Abelian
$\msu(n)$ sector, but is reconciled with 
the mild running of the non-Abelian sector above the non-commutativity scale.
This suggests that taking into account the full RG flow, 
the strong sensitivity of the cosmological constant on the 
energy scale in the IR becomes mild at $\L_{NC}$.
A similar statement applies to the quadratically divergent  
Einstein-Hilbert term.
It is precisely this behavior which should reconcile the 
apparent non-renormalizability of gravity with the good
renormalization behavior of Yang-Mills gauge theory, 
which in the case of the IKKT  model (or closely related models) is
expected to be finite.

More specifically,
recall that the bare cosmological constant
is given by the $\mmu(1)$ sector
of the Yang-Mills term $-(2\pi)^2 \Tr [X^a,X^b][X_a,X_b] 
\sim 4\int d^4x\sqrt{g}$ in the matrix model\footnote{
assuming $g_{\mu\nu} = G_{\mu\nu}$; however additional
higher-order terms in the action may also contribute.}.
It is thus quite conceivable that taking into account quantum corrections,
the cosmological constant
is small in the IR, but merges with the 
$\msu(n)$ Yang-Mills action at $\L_{NC}$.  
However while it is consistent, 
this does not yet explain why the vacuum energy 
should indeed be small in the IR. 
At this point, it is interesting to recall that 
flat $\R^4$  (in fact any harmonically embedded space) is a solution also in presence
of an arbitrarily large vacuum energy in the matrix model,
unlike in General Relativity. It remains to be seen
if this observation carries over in some way to (modified)
solutions of Einstein-Hilbert type.

\section{Conclusion}\label{sec:conclusion}

We have shown how gravitational actions including the Einstein-Hilbert action
(with an additional scale factor) can be obtained as higher-order terms
in matrix models of Yang-Mills type. The resulting actions 
are consistent with the induced gravitational 
terms in the quantum effective action of the matrix model, as obtained
previously using a semi-classical heat kernel computation \cite{Klammer:2009dj}. 
This exhibits the gravity sector in 
these matrix models more explicitly, and allows to identify
and control the precise form of the gravitational action. 
In general, both the extrinsic and the intrinsic geometry
of the space-time brane $\cM^4 \subset \R^D$ play a role. However, for 
special cases only the intrinsic geometry enters, as in General Relativity.
This allows a controlled study of the gravitational sector 
of the matrix model at the quantum level. It provides 
a new, non-perturbative and background-independent approach to quantum gravity
where space and geometry emerge at low energies but are not put in by hand.

There are  different avenues which can be pursued further. First, one
can focus on the pure gravity sector of the (bosonic) matrix model using the 
additional terms introduced in this paper. At the 
quantum level, this would essentially require the systematic study of 
a suitable version of the renormalization group flow for this type of 
matrix model, e.g. by scaling the size of the matrices, or 
using an explicit cutoff term such as $\Box X^a\Box X_a$.
Alternatively, one can consider finite versions of the matrix model, 
notably the IKKT model \cite{Ishibashi:1996xs}, and closely related models such as 
\cite{Chatzistavrakidis:2010xi}.
In that case, the gravitational action emerges as part or
the quantum effective action in a finite ({\nc}) model
including gauge fields and matter. One  
may hope that this leads to a fully consistent quantum theory of all
fundamental interactions including gravity.
However, much more work is required before such a long-term goal can be achieved.

\subsection*{Acknowledgements}

Useful discussions with A. Schenkel are gratefully acknowledged.
This work was supported by the ``Fonds zur F\"orderung der Wissenschaftlichen Forschung'' (FWF) under contract P21610-N16.

\startappendix
\Appendix{Derivation of \eq{S6-geom}}\label{sec:appendix}
%
We assume $G_{\mu\nu} = g_{\mu\nu}$ throughout.
The first part of this result follows immediately from the following 
identity \eq{box-laplace-id}
\begin{align}
\Box X^a &\sim -\pb{x^b}{\pb{x^{c}}{x^a}} \eta_{bc}
= -e^\sigma \lap x^a 
\end{align}
together with $\rho= \sqrt{g}e^{-\sigma}$.
The second term is more difficult to analyze. 
We use the (constant) background metric $\eta_{ab}$ to pull down Latin indices, i.e. $x_a\equiv x^b\eta_{ab}$, and consider first
\bea
&& (2\pi)^2\Tr [X^c,[X^a,X^b]] [X_c,[X_a,X_b]] 
\sim \intg  \del_\rho\pb{x^a}{x^b} \del^\rho\pb{x_a}{x_b} \nn\\
&=&  \intg  
\nabla_\rho (\theta^{\a\b}\del_\a x^a \del_\b x^b) 
\nabla^{\rho} (\theta^{\mu\nu}\del_\mu x_a \del_\nu x_b) \nn\\
&=&  \intg  \Big(
\nabla_\rho \theta^{\a\b}\nabla^{\rho} \theta^{\mu\nu}
\del_\a x^a \del_\b x^b \del_\mu x_a \del_\nu x_b 
+ 2 \theta^{\mu\nu}\nabla_\rho \theta^{\a\b}
\del_\a x^a \del_\b x^b \nabla^{\rho} (\del_\mu x_a \del_\nu x_b) \nn\\
&& + \theta^{\a\b}\theta^{\mu\nu}
\nabla_\rho (\del_\a x^a \del_\b x^b) 
\nabla^{\rho} (\del_\mu x_a \del_\nu x_b) \Big)\nn\\
&=&  \intg  \Big(
\nabla_\rho \theta^{\a\b}\nabla^{\rho} \theta^{\mu\nu}
 g_{\a\mu} g_{\b\nu}
+ 4 \theta^{\mu\nu}\nabla_\rho \theta^{\a\b}
\del_\a x^a \del_\b x^b  \del_\mu x_a \nabla^{\rho}\del_\nu x_b \nn\\
&& + \theta^{\a\b}\theta^{\mu\nu}\(
2\nabla_\rho \del_\a x^a \del_\b x^b
\nabla^{\rho} \del_\mu x_a \del_\nu x_b
+ 2\nabla_\rho \del_\a x^a \del_\b x^b
\del_\mu x_a \nabla^{\rho} \del_\nu x_b
\)\Big) . 
\eea
Using \eq{3-nabla-vanish} as well as
\be
 g_{\mu\mu'}g_{\b\b'} \theta^{\mu'\b'} 
= -e^{\sigma} \theta^{-1}_{\mu\b},
\ee
this simplifies as
\begin{align}
& (2\pi)^2\Tr [X^c,[X^a,X^b]] [X_c,[X_a,X_b]] \nn\\
&\sim  \intg  \Big(
-\nabla_\rho \theta^{\a\b}\nabla^{\rho} (e^\sigma\theta^{-1}_{\a\b})
 + 2\theta^{\a\b}\theta^{\mu\nu}
\nabla_\rho \del_\a x^a \nabla^{\rho} \del_\mu x_a g_{\b\nu}\Big)  \nn\\
&=  \intg  \Big(
-\nabla_\rho \theta^{\a\b}\theta^{-1}_{\a\b}\del^{\rho} e^\sigma
-e^\sigma\nabla_\rho \theta^{\a\b}\nabla^{\rho} \theta^{-1}_{\a\b}
 + 2 e^\sigma g^{\a\mu}
(-R_{\a\mu} + \lap x^a \nabla_\a\del_\mu x_a)\Big)  \nn\\
&=  \intg  \Big(
 2e^\sigma \del^{\mu}\sigma \del_\mu\sigma 
-e^\sigma\nabla_\rho \theta^{\a\b}\nabla^{\rho} \theta^{-1}_{\a\b}
 + 2 e^\sigma (-R+ \lap x^a \lap x_a)\Big)
 \nn\\
&= \intg e^\sigma \Big(
e^{-\sigma}\theta^{\mu\rho} \theta^{\eta\a} R_{\mu \rho\eta\a} 
 - 4 R
 + 4\del^{\mu}\sigma \del_\mu\sigma 
 + 2 \lap x^a \lap x_a\Big)
, \nn
\end{align}
where the identities of Lemma~\ref{lem:app-identities1} were used. 
Hence, we get
\bea
&& (2\pi)^2\Tr [X^c,[X^a,X^b]] [X_c,[X_a,X_b]] \nn\\
&\sim& 2\intg  \Big(
\theta^{\rho\a} \theta^{\mu\eta} R_{\mu \rho\eta\a} 
 - 2 e^\s R 
 + 2 e^\s \del^{\mu}\s \del_\mu\s 
 + e^\sigma \lap x^a \lap x_a)\Big) 
\,.
\eea

\begin{lemma}
\label{lem:app-identities1}
The following identities are useful:
\begin{subequations}
\begin{align}
\nabla_\mu(e^{-\sigma}\theta^{\mu\nu}) &= 0,
\label{del-theta-id}  \\
\pb{x^b}{\pb{x^{c}}{x^a}} \eta_{bc}
&= e^\sigma \Box_G x^a \, ,
\label{box-laplace-id} \\
\intg e^\sigma  g^{\rho\rho'} \nabla_\rho\theta^{-1}_{\mu\a}
\nabla_{\rho'}\theta^{\a\mu}
&=  \intg  e^\sigma 
\left(e^{-\sigma}\theta^{\mu\rho} \theta^{\eta\a} R_{\mu \rho\eta\a} - 2R
 + 2\del^{\mu}\sigma \del_\mu\sigma \right) ,
\label{deltheta-deltheta-id} \\
R_{\l\mu\nu\rho}\theta^{\l\mu}\theta^{\nu\rho}
&= 2R_{\l\nu\mu\rho}\theta^{\l\mu}\theta^{\nu\rho} \, ,
\label{Rthth-id-2} \\
\theta^{-1}_{\mu\nu} \nabla_\a \theta^{\mu\nu} 
&= - 2\del_\a \sigma  \,.
\label{thetadiv}
\end{align}
\end{subequations}

\end{lemma}
\begin{proof}
\eq{del-theta-id} and \eq{box-laplace-id}
was shown in \cite{Steinacker:2008ya}.
To show \eq{deltheta-deltheta-id}, we 
use the Jacobi identity and proceed as follows:
\begin{align}
\intg e^\sigma  g^{\rho\rho'} \nabla_\rho\theta^{-1}_{\mu\a}
\nabla_{\rho'}\theta^{\a\mu}
&= - \intg  e^\sigma g^{\rho\rho'} 
(\nabla_\mu\theta^{-1}_{\a\rho} + \nabla_\a\theta^{-1}_{\rho\mu})
\nabla_{\rho'}\theta^{\a\mu}  \nn\\
&= - 2\intg e^\sigma  g^{\rho\rho'} 
\nabla_\mu\theta^{-1}_{\a\rho} \nabla_{\rho'}\theta^{\a\mu}  \nn\\
&= 2 \intg  e^\sigma g^{\rho\rho'} \Big(
\theta^{-1}_{\a\rho} \nabla_\mu\nabla_{\rho'}\theta^{\a\mu} 
 + \theta^{-1}_{\a\rho} \nabla_{\rho'}\theta^{\a\mu} \del_\mu\sigma \Big) \nn\\
&=  2\intg   e^\sigma  (g^{\rho\rho'} 
\theta^{-1}_{\a\rho}[\nabla_\mu,\nabla_{\rho'}]\theta^{\a\mu} 
+\del^{\mu}\sigma \del_\mu\sigma)\nn\\
&=  2\intg  e^\sigma\!\(  g^{\rho\rho'} \theta^{-1}_{\a\rho}
(-{R_{\mu \rho'\eta}}^\a\theta^{\eta\mu} 
 -{R_{\mu \rho'\eta}}^\mu\theta^{\a\eta} )
 +\del^{\mu}\sigma \del_\mu\sigma\)\nn\\
&=  2\intg  e^\sigma 
\(e^{-\sigma}\theta^{\a\rho} \theta^{\eta\mu} R_{\mu\rho\eta\a} - R
 + \del^{\mu}\sigma \del_\mu\sigma \)
\end{align}
using \eq{del-theta-id} i.e.
$\nabla_\mu\theta^{\mu\nu} =  \theta^{\mu\nu}\del_\mu \sigma$,
\eq{Rthth-id-2}  as well as 
\be
\nabla^{\rho}\theta^{-1}_{\a\rho} =0
\ee
which holds for $G=g$.
Finally \eq{thetadiv} follows from the fact that 
\be
{\cJ^\mu}_\nu := e^{-\sigma/2} \theta^{\mu\mu'} g_{\mu'\nu}
\ee
is unimodular $\det \cJ = 1$, which implies
\bea
0 &=& \del_\a \det \cJ = {(\cJ^{-1})^\mu}_\nu \nabla_\a {\cJ^\nu}_\mu \nn\\
&=&  e^{2\sigma}\del_\a e^{-2\sigma} +
g^{\mu\s}\theta^{-1}_{\s\nu} \nabla_\a \theta^{\nu\eta}g_{\eta\mu} \nn\\
&=&  -2 \del_\a \sigma + \theta^{-1}_{\eta\nu} \nabla_\a \theta^{\nu\eta}.
\eea

\end{proof}



\end{document}